\documentclass[aps,prc,reprint,superscriptaddress,nofootinbib]{revtex4-1}
\usepackage{graphicx}
\usepackage{multirow}
\usepackage{comment}
\usepackage{color}
\usepackage{amssymb}
\usepackage{amsmath}
\usepackage{lineno}
\newcommand{\psixprior}{\psi^\mathrm{prior}_x} 
\newcommand{\psixpost}{\psi^\mathrm{post}_x} 
\newcommand{\psixno}{\psi^\mathrm{NO}_x}

\def\nuc#1#2{\relax\ifmmode{}^{#1}{\protect\text{#2}}\else${}^{#1}$#2\fi}

\newcommand{\vecr}{{\vec r}}

\newcommand{\be}{\begin{eqnarray}}
\newcommand{\ee}{\end{eqnarray}}

\newcommand{\bwt}{\begin{widetext}}
\newcommand{\ewt}{\end{widetext}}

\begin{document}
%\begin{CJK*}{GB}{song}
%\preprint{}

\title{Numerical assessment of post-prior equivalence for inclusive breakup reactions}


\author{Jin Lei}
\email[]{jinlei@us.es}
%\thanks{}

\affiliation{Departamento de FAMN, Universidad de Sevilla, 
Apartado 1065, 41080 Sevilla, Spain.}

\author{Antonio M. Moro}
\email[]{moro@us.es}
%\thanks{}

\affiliation{Departamento de FAMN, Universidad de Sevilla, 
Apartado 1065, 41080 Sevilla, Spain.}


\begin{abstract}
We address the problem of the post-prior equivalence in inclusive breakup reactions 
induced by weakly-bound nuclei. The problem is studied within the DWBA model of Ichimura, Austern, Vincent [Phys. Rev. C32, 431 (1985)]. 
The post and prior formulas obtained in this model are briefly recalled, and applied to 
several breakup reactions induced by deuterons and $^6$Li projectiles, to test their 
actual numerical equivalence.  The different contributions of the prior-form formula are also discussed. A critical comparison with the prior-form distorted-wave Born approximation (DWBA) model of 
Udagawa and Tamura [Phys. Rev. C24, 1348 (1981)] is also provided. 
\end{abstract}

% 25.70.Mn, Projectile and target fragmentation
% 24.10.Eq 	Coupled-channel and distorted-wave models
% 25.45.-z  2H-induced reactions
% 24.87.+y 	Surrogate reactionsç

%\keywords{}
\date{\today}%
\maketitle

\section{Introduction \label{sec:intro}}
%-------------------------------------------------

Breakup is an important mechanism occurring in nuclear reactions involving weakly-bound nuclei. The analysis of these reactions has provided useful information on the structure of weakly-bound nuclei (such as separation energies, angular momenta/parities, electric responses to the continuum, etc). A  detailed understanding of these  processes is also necessary for a number of applications, such as in $(d,pf)$ {\it surrogate} reactions \cite{Esc12} or  the production of radioisotopes for medical purposes \cite{Bet11}.

%When the projectile nucleus is broken up into two fragments, 
For two-body projectiles, these reactions can be represented as $a+A \rightarrow b+ x+ A$, where $a=b+x$. If the three outgoing particles are observed in a definite final state the reaction is said to be {\it exclusive}. This problem can be treated as an effective scattering problem with three particles interacting via some effective two-body interactions. Although the exact, rigorous solution of this problem  can in principle be obtained  solving the so-called Faddeev equations \cite{Fad60}, in practice the complexity of this method limits so far its applicability to specific situations. For this reason, alternative approaches, such as the popular continuum-discretized coupled-channels (CDCC) method \cite{Aus87}, have been used.  At higher energies, semiclassical approaches become an efficient and appealing alternative (e.g., \cite{Typ94,Esb96}). 
%\cite{Typ94,Esb96,Kid94,Cap04}.
 
%However, in many situations, the final state is not fully specified. 
%A number of theories have been developed for these processes, for example, the continuum-discretized coupled-channels (CDCC) method, the Faddeev 

A qualitatively different scenario occurs when  the final state is not fully specified. For example, this is the case of reactions of the form $A(a,bX)$, in which only one of the projectile constituents (say, $b$) is observed. In this case the reaction is said to be {\it inclusive} with respect to the unobserved particle(s). The simplest process contributing to the inclusive cross section is that in which the three outgoing particles remain in their ground states, which receives the name of {\it elastic breakup} (EBU). However, more complicated processes are possible, for example, breakup accompanied by $x$ or $A$ excitation,  by particle transfer between $x$ and $A$, or by fusion of $x$ with $A$ (incomplete fusion, ICF). The sum of these contributions is referred to as non-elastic breakup (NEB). 

%\sout{This includes the EBU channel, in which $x$ and $A$ remain in their ground state, but also $v$ transfer, breakup accompanied by excitations of $A$, and  $v$+$A$ fusion [named {\it incomplete fusion} (ICF)].   These non-elastic breakup components (NEB) must be added to the  EBU component to give the total inclusive breakup. }

Due to the large number of accessible states, a detailed calculation of the NEB part, in which all these processes are included explicitly, is in general not possible. For that reason, in the 1980s several groups developed closed-form expressions in which the sum over final states was done in a formal way, using completeness of the $x+A$ states \cite{Bud78,Bau80,Shy80,Uda81,Uda84,Aus81,Kas82,HM85,Ich85}.  
%Whereas the calculation of EBU can be accurately done within the CDCC method and other approaches, the calculation of NEB is more involved due to the large number of accessible states. The problem was addressed in the 1980s by several groups. The common aspect of these models is that they provide a procedure to sum over final of the $x+A$ system, without introducing these states explicitly.
%{\bf REDUCE/ELIMINATE?: In the pioneering works by Baur and co-workes  \cite{Bud78,Bau80,Shy80}, the sum was done making use of unitarity and a surface approximation of the form factors of excited states of the  residual nucleus.  These two approximations were avoided in later works  by Udagawa and Tamura \cite{Uda81,Uda84}, who used a prior-form DWBA  formalism, and by Austern and Vincent (UT) \cite{Aus81}, who  used the post-form  DWBA. The latter was refined by Kasano and Ichimura \cite{Kas82},  who   found a formal separation between the EBU and NEB contributions. These  results were carefully reviewed by Ichimura, Austern and Vincent (AV) \cite{Ich85} and the model was subsequently referred to as the IAV formalism.}
Here, we focus  on the models proposed by Udagawa and Tamura (UT hereafter) \cite{Uda81,Uda84} and by Ichimura, Austern and Vincent (IAV hereafter) \cite{Aus81,Kas82,Ich85}. The main difference between these models is  that, whereas  UT use the prior-form DWBA, IAV employ the post-form representation. Although the final expressions for these models have the same formal structure (see Sec.~\ref{sec:formalism}), they lead to different predictions for the NEB cross sections. This is in contrast to the DWBA formula for transfer between bound states, where it is well known that the post and prior formulas are fully equivalent.  This discrepancy led to a long-standing controversy between these two groups, which lasted for more than a decade. At the heart of the discussion was the fact that the transformation of the post form DWBA expression of IAV to its prior form gave rise to additional terms, not present in the UT prior formula. These additional terms guaranteed the  post-prior equivalence for NEB, but they were nevertheless regarded as unphysical by UT. To support their conclusions, UT performed calculations for several inclusive reactions \cite{Li84,Mas89}, in which they showed that the IAV calculations largely overestimated the data.  

The IAV model has been recently revisited and implemented by several groups \cite{Car15,Pot15,Jin15}. Contrary to the referred results of Udagawa, Tamura and collaborators, the comparison of these recent calculations with available data has shown very encouraging results.  These calculations have been performed using either the original post-form formulation \cite{Car15,Jin15} or its (in principle) equivalent prior form \cite{Pot15}. However, a consistent comparison between the post and prior results has not been made to our knowledge.  One of the reasons is that a direct evaluation of the post-form formula is not feasible, owing to the marginal convergence of the post-form breakup amplitudes. To overcome this problem, several regularization procedures have been suggested, such as the integration in the complex plane of Vincent and Fortune \cite{Vin70}, the introduction of a convergence damping factor \cite{Hub65,Vin68} or the replacement of the oscillatory distorted waves of the outgoing $b$ fragments by some averaged wave packets \cite{Tho11}. The convergence and stability of these procedures need to be carefully examined.  

% w. The  work of IAV \cite{Ich85} and subsequent works by other authors (see, e.g., \cite{Hus90}) served to clarify the situation, and to evidence that these additional terms are essential to preserve the important post-prior equivalence. For this reason, recent implementations have made use of the DWBA model of IAV, either in its original post-form \cite{Car15,Jin15} or in its in principle equivalent  prior-form \cite{Mas89,Pot15}. Although the comparison of these models with the data have shown very encouraging results, the post-prior equivalence has never been tested in practice, to our knowledge. One of the reasons is that a direct evaluation of the post-form formula is not feasible owing to the marginal convergence of the post-form breakup amplitudes. Several regularization procedures have been suggested nevertheless to  circumvent this problem, such as the introduction of a convergence damping factor \cite{Hub65,Vin68} or the replacement of the oscillatory distorted waves of the outgoing $b$ fragments by an average wave packet \cite{Tho11}. The convergence and stability of these procedures needs to be carefully examined.  

The  goal of this work is manifold. First, we aim to assess,  in a quantitative way, the actual equivalence of the post and prior NEB formulas of the IAV model. For that, we will apply these formulas to specific cases.  Furthermore, this study will serve to test the validity of the regularization procedure of the post-form integrals invoked in \cite{Tho11,Jin15}. In each case, we compare also with the UT model and with available data, in order to assess the validity of these models against the data. Finally, we aim at examining the relative importance of the different terms entering the prior-form expression. For that, we have performed calculations for $^{62}$Ni($d$,$p$X) at  $E=25.5$~MeV and  $^{209}$Bi($^6$Li,$\alpha$X) at $E=36$~MeV.

%The motivation is manyfold: first, this will allow to assess the {\it binning} procedure used in our previous work; second, it will permit to evaluate the relative importance of the different terms appearing in the prior form and, finally, by comparing with available experimental data it will allow to judge on the validity of the IAV and UT models.  

The paper is organized as follows. In Sec.~\ref{sec:formalism} we summarize the main formulas of the IAV and UT models, and outline the relation between them. In Sec.~\ref{sec:calc}, the 
formalism is applied to several inclusive reactions induced by deuterons and $^{6}$Li. Finally, in Sec.~\ref{sec:sum} we summarize the main results.

%---------------------------------------------------------------------------
\section{\label{sec:formalism} Post and prior formulas for inclusive breakup}
%----------------------------------------------------------------------------
In this section we briefly review the  main results of the UT and IAV models. Further details can be found in the referred works as well as in our preceding paper \cite{Jin15}. We write the process under study as
\begin{equation}
a (=b+x) + A \rightarrow b + B^* .
\end{equation}
%in which a projectile $a$ colliding with a target $A$ dissociates into two fragments $b$ and $x$. 
We assume that the experiment is inclusive with respect to the particle $x$. Consequently, only $b$ is observed and the corresponding experimental cross sections will correspond to a sum over all possible final states of the $x+A$ system. This includes the EBU as well as the NEB components mentioned in the introduction. 

The IAV model, as well as the UT model, treats the $b$ particle as an {\it spectator}, meaning that its interaction with the target nucleus is described with an optical potential $U_{bA}$.

Using the  post-form DWBA, the inclusive breakup differential cross section, as a function of the detected angle and energy of the fragment $b$, is given by
\begin{widetext}
\begin{equation}
\frac{d^2\sigma}{d\Omega_b E_b } = \frac{2 \pi}{\hbar v_a} \rho(E_b) \sum_{c} |\langle \chi^{(-)}_{b} \Psi^{c,(-)}_{xA} |V_\mathrm{post}| \chi_a^{(+)} \phi_a  \phi^{0}_A \rangle |^2 \delta(E-E_b-E^c) ,
\end{equation}
\end{widetext}
where $V_\mathrm{post} \equiv V_{bx} + U_{bA}-U_{bB}$ is the post-form transition operator%
\footnote{In their original papers \cite{Ich85}, IAV usually make the approximation $V_\mathrm{post}\approx V_{bx}$, thus neglecting the so-called remnant term, $U_{bA}-U_{bB}$. In Ref.~\cite{Jin15} we showed that this is a good approximation for deuterons on heavy targets, but not for $^{6}$Li reactions. In this work we retain the full transition operator, for an accurate comparison between the post and prior results.}, 
 $\rho_b(E_b)=k_b \mu_{b} /((2\pi)^3\hbar^2)$ (with $\mu_b$ the reduced mass of $b+B$ and $k_b$ their relative wave number) , $\phi_a(\vec{r}_{bx})$ and $\phi^{0}_A$ are the projectile and target ground-state wave functions,  $\chi_a^{(+)}$ and $\chi^{(-)}_{b}$  are distorted waves describing the $a-A$ and $b-B$ relative motion, respectively, and $\Psi^{c,(-)}_{xA}$ are the eigenstates of the $x+A$ system, with $c=0$ denoting the $x$ and $A$ ground states. Thus, for $c=0$ this expression gives the EBU part, whereas the terms $c \neq 0$ give the NEB contribution.   

The theory of IAV allows to perform the sum in a formal way, making use of the Feshbach projection formalism and the optical model reduction, leading to a closed form for the NEB differential cross section:
\begin{equation}
\label{eq:iav}
\left . \frac{d^2\sigma}{dE_b d\Omega_b} \right |^\mathrm{IAV}_\mathrm{NEB} = -\frac{2}{\hbar v_{i}} \rho_b(E_b) 
 \langle \psixpost | W_x | \psixpost \rangle   ,
\end{equation}
where $W_x$ is the imaginary part of the optical potential $U_x$, which describes $x+A$ elastic scattering.  The function  $\psi_x^\mathrm{post}(\vecr_x)$  (the $x$-channel wave function hereafter) describes the $x-A$ relative motion when the target is in the ground state and the $b$ particle scatters with momentum $\vec{k}_b$, and is obtained by solving the inhomogeneous equation
%in the state $\chi_b^{(-)}(\vec{k}_b, \vec{r}_b$), corresponding to a given energy and scattering angle. 
%
\begin{equation}
\label{phix_post}
(E^+_x - K_x - {U}_x)  \psi_x^\mathrm{post}(\vecr_x) =  (\chi_b^{(-)}| V_\mathrm{post}|\phi_a  
\chi^{(+)}_{a}\rangle .
\end{equation}
where $E_x=E-E_b$. %, and $V_\mathrm{post} \equiv V_{bx}+U_{bA}-U_{b}$ is the post-form transition operator. 

%and  ${U}_x$ the formal optical model potential describing $x$-$A$ elastic scattering. 

Note that the result (\ref{eq:iav}) bears some resemblance with the well-known optical theorem, which provides the total reaction (absorption) cross section in two-body scattering. This analogy was in fact exploited in Ref.~\cite{Jin15} to derive Eq.~(\ref{eq:iav}), using a {\it generalized optical theorem} \cite{Cot10}.

Udagawa and Tamura \cite{Uda81} derived a very similar formula for the same problem, but making use of the prior form DWBA. Their final result is formally identical to 
Eq.~(\ref{eq:iav}), but with the $x$-channel wave function given by $\psixprior$, which is a solution of 
%
\begin{comment} 
\begin{equation}
\label{eq:ut}
\left . \frac{d^2\sigma}{dE_b d\Omega_b} \right |^\mathrm{UT}_\mathrm{NEB} = -\frac{2}{\hbar v_{i}} \rho_b(E_b) 
 \langle \psixprior | W_x | \psixprior \rangle   ,
\end{equation}
where the $x$-channel wave function $\psi_x^\mathrm{prior}(\vecr_x)$ is obtained from 
\end{comment}
%
\begin{equation}
\label{phix_prior}
(E^+_x - K_x - {U}_x)  \psi_x^\mathrm{prior}(\vecr_x) =  (\chi_b^{(-)}| V_\mathrm{prior}| \chi^{(+)}_{a} \phi_a \rangle ,
\end{equation}
with $V_\mathrm{prior} \equiv U_{xA} + U_{bA}-U_{aA}$. 

Despite their formal analogy, the UT and IAV expressions lead to different predictions for the NEB cross sections.
% This difference was the origin of a lengthly discussion/controversy between the IAV and UT groups. These works, along with some other works by Hussein and collaborators \cite{Hus90}  helped to understand the connection between these two models.
 An important  result to understand the connection between these two expressions is the relation \cite{Li84}
\begin{equation}
\label{eq:psipopr}
\psixpost = \psixprior +  \psixno ,
\end{equation}
where 
\begin{equation}
\label{eq:psino}
%\psixno =  \langle \chi^{(-)}_b |  \chi_a^{(+)} \phi_a (\vec{r}_{bx})  \rangle
\psixno(\vec{r}_x) =  \langle \chi^{(-)}_b |  \chi_a^{(+)} \phi_a  \rangle ,
\end{equation}
is the so-called non-orthogonality NO overlap. %This relation was first derived by Li, Udagawa and Tamura \cite{Li84}.

% 1column
%\begin{align}
%\label{eq:post_prior}
%\left . \frac{d^2\sigma}{dE_b d\Omega_b} \right |^\mathrm{post}_\mathrm{NEB} & = 
%\left . \frac{d^2\sigma}{dE_b d\Omega_b} \right |^\mathrm{UT}_\mathrm{NEB} 
% + \left . \frac{d^2\sigma}{dE_b d\Omega_b} \right |^\mathrm{NO}_\mathrm{NEB}
% + \left . \frac{d^2\sigma}{dE_b d\Omega_b} \right |^\mathrm{IN}_\mathrm{NEB} ,
%\end{align}

Replacing (\ref{eq:psino}) into Eq.~(\ref{eq:iav}) one gets 
\begin{widetext}
\begin{align}
\label{eq:post_prior}
\left . \frac{d^2\sigma}{dE_b d\Omega_b} \right |^\mathrm{IAV}_\mathrm{NEB} & = 
\left . \frac{d^2\sigma}{dE_b d\Omega_b} \right |^\mathrm{UT}_\mathrm{NEB} + 
\left . \frac{d^2\sigma}{dE_b d\Omega_b} \right |^\mathrm{NO}_\mathrm{NEB} +
\left . \frac{d^2\sigma}{dE_b d\Omega_b} \right |^\mathrm{IN}_\mathrm{NEB} ,
\end{align}
\end{widetext}
where we have introduced the  non-orthogonality (NO)  {\it cross section}
\begin{equation}
\label{eq:no}
\left . \frac{d^2\sigma}{dE_b d\Omega_b} \right |^\mathrm{NO}_\mathrm{NEB} = 
-\frac{2}{\hbar v_{i}} \rho_b(E_b) 
 \langle \psixno | W_x | \psixno \rangle ,
\end{equation}
and the interference (IN) term
\begin{equation}
\label{eq:int}
\left . \frac{d^2\sigma}{dE_b d\Omega_b} \right |^\mathrm{IN}_\mathrm{NEB} = 
- \frac{4}{\hbar v_a} \rho_b(E_b) 
 \mathrm{Re}  \langle \psixprior | W_{xA} | \psixno \rangle .
\end{equation}

Equation (\ref{eq:post_prior}) represents the post-prior equivalence of the NEB cross sections in the IAV model, with the RHS corresponding to the prior-form expression of this model. The first term is just the UT formula, which is formally analogous to the IAV post-form formula (\ref{eq:iav}), but with the $x$-channel wave function given by $\psi_x^\mathrm{prior}(\vecr_x)$. The two additional terms, which are responsible for the discrepancy of the IAV and UT results,   arise from the NO overlap. These terms ensure the post-prior equivalence of the NEB cross sections. However, UT considered that these two additional terms are unphysical and hence that the post-prior equivalence does not hold for the NEB. We note here that this problem does not arise for the EBU part, for which the post and prior formulas are well known to give identical results \cite{Ich85}. 
%According to IAV, the NEB cross section must contain the three terms of Eq.~(\ref{eq:post_por}). This, in turn, guarantees the post-prior equivalence.   On the contrary, UT considered that the correct formula for NEB is given by Eq.~(\ref{eq:ut}), and that the NO overlap $\psixno$, which is responsible of the second and third terms in the prior-form IAV formula, Eq.~(\ref{eq:post_prior}), is unphysical.  
To support their interpretation, Mastroleo, Udagawa and Tamura \cite{Mas89} performed calculations for  the reactions $^{58}$Ni($\alpha$,$pX$) at $E_\alpha=80$~MeV and $^{62}$Ni($d$,$pX$) at $E_d=25.5$~MeV.
%One of the few available calculations using Eq.~(\ref{eq:post_prior}) was due Mastroleo, Udagawa and Tamura \cite{Mas89}, who applied this formula to the reactions $^{58}$Ni($\alpha$,$pX$) at $E_\alpha=80$~MeV and $^{62}$Ni($d$,$pX$) at $E_d=25.5$~MeV.
 In both cases, they found that the sum of the EBU (calculated with DWBA) and the NEB (calculated with the IAV model) overestimates the data. This result was interpreted as  evidence for the failure of the IAV model, and  support for the UT theory. 

This interpretation was later questioned in subsequent works by Ichimura {\it et al.}~\cite{Ich86,Ich88,Ich90} and also by Hussein and co-workers \cite{Hus90}. These works clearly demonstrated that the UT formula provides only the so-called {\it elastic breakup fusion} component, which corresponds to breakup without simultaneous excitation of the target $A$ by the interaction $V_{xA}$, and that the prior-post equivalence does indeed hold for inclusive processes as well. 

Despite these intense formal developments, the post-prior equivalence for NEB, represented by Eq.~(\ref{eq:post_prior}), has never been numerically tested to our knowledge. One of the reasons is %the marginal convergence of the post-form formula mentioned in the introduction. 
that the direct solution of Eq.~(\ref{phix_post}) is not possible due to the oscillatory behavior of the source term. This, in turn, is a consequence of the oscillatory behavior of the scattering wave function $\chi_b^{(-)}$, which is not damped asymptotically by either the initial state wave function $\phi_a$ or by the transition operator $V_\mathrm{post}$. Notice that this problem does not arise in the prior form because, in this case, the transition operator ($V_\mathrm{prior}$) makes the source term short-ranged. As noted in the introduction, some regularization procedures have been proposed in the literature to overcome this problem.  Here, we adopt the method proposed in Ref.~\cite{Tho11}, which consists in averaging  the distorted waves $\chi_b^{(-)}$ over small momentum intervals ({\it bins}). The resulting averaged functions become square-integrable and the source term of Eq.~(\ref{phix_post})  vanishes at large distances. This procedure was successfully applied in our previous work \cite{Jin15} to several reactions.

%A way to overcome the marginal convergence of the post-form formula is to introduce some averaging procedure of the  $\chi_b^{(-)}$ distorted waves, by integrating them over finite energy or momentum intervals. The resulting averaged functions become square-integrable and the source term of Eq.~(\ref{phix_post}) will vanish at large distances. This procedure was proposed in Ref.~\cite{Tho11} and successfully applied in our previous work \cite{Jin15} to several reactions. 
% It is the purpose of the present work to fully test the equivalence (\ref{eq:post_prior}) in some specific reactions. This, in turn, will serve to validate the regularization procedure of the post-form formula used in \cite{Jin15}. In addition, we have reexamined the $^{62}$Ni($d$,$pX$) case analyzed in Ref.~\cite{Mas89} to verify the conclusions of UT.  

In the following section, we apply the IAV and UT models to specific reactions comparing, in the former, the prior and post results.

%------------------------------------------------------------------
\section{\label{sec:calc} Calculations}
%------------------------------------------------------------------
As a first example, we consider the reaction $^{62}$Ni($d$,$pX$) at 
$E_d$=25.5~MeV, which will allow to compare our results with those from
Ref.~\cite{Mas89}.  

In our calculations, the deuteron ground-state wavefunction was generated with the simple Gaussian 
potential of Ref.~\cite{Aus87}. The deuteron and proton distorted waves are generated with the same 
optical potentials used in  Ref.~\cite{Mas89}. 
As noted in the previous section, to evaluate the post-form formula the distorted waves  $\chi_b$ are averaged over small momentum intervals. 
%the post-form formula cannot be evaluated  directly, owing to the slow convergence of the integrals. For that reason,  a binning procedure, similar to that employed in Refs.~\cite{Tho11,Jin15},  is used, consisting in averaging the distorted waves $\chi_b$ over small  momentum bins. This procedure produces square-integrable functions, which make the integrals converging.
 Although this procedure is not required for the 
prior-form formula, to have consistent ingredients in both calculations, 
the same averaged  distorted waves were used in that case.

%---------------------------------------------------------------------------------------------------
% 62Ni(d,p)
%----------------------------------------------------------------------------------------------------
\begin{figure}[tb]
\includegraphics[width=0.85\columnwidth]{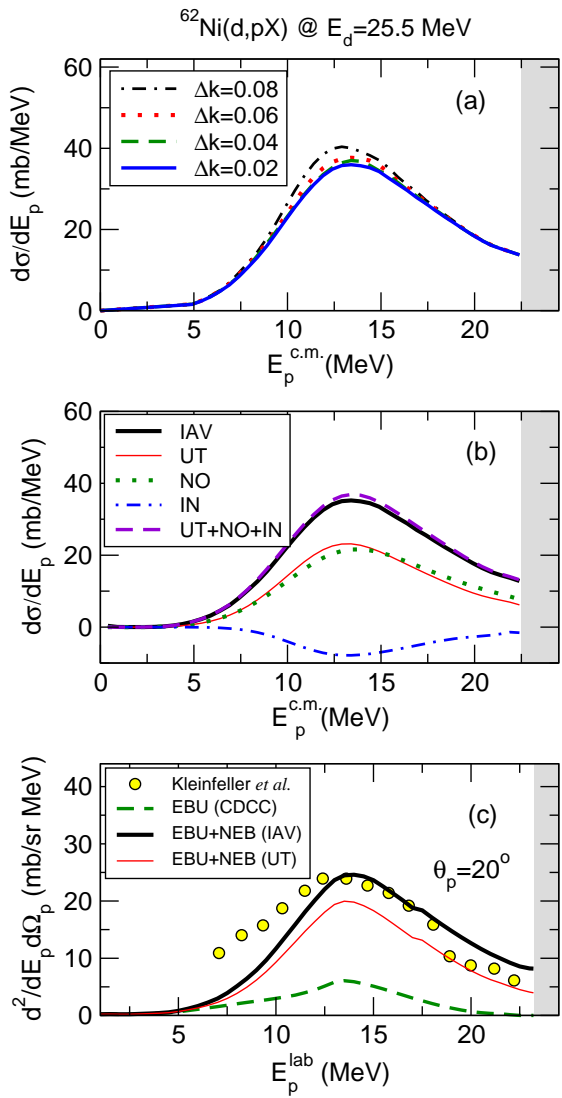}% Here is how to import EPS art
\caption{\label{ni62dp_dsde} (Color online) Proton energy spectra for $^{62}$Ni($d$,$p$X) at $E_d$=25.5~MeV (a) Convergence of the angle-integrated energy distribution (in c.m.) of the NEB cross sections with respect to the bin width (in fm$^{-1}$) used for the $b$ distorted waves. (b) Comparison of the post and prior results. 
(c) Comparison of IAV and UT models with the data from Refs.~\cite{Kle81,Mas89}, corresponding to the double differential cross section, as a function of the proton energy in LAB frame, for $\theta_p=20^\circ$.}
\end{figure}

%\begin{figure}[tb]
%\begin{center}
% {\centering \resizebox*{0.95\columnwidth}{!}{\includegraphics{ni62dp_dsde.eps}} \par}
%\caption{\label{ni62dp_dsde}(Color online) .}
%\end{center}
%\end{figure}
%-------------------------------------------------------------

Before comparing the post and prior results, we investigate the convergence of the post-form formula with respect to 
the bin size,  $\Delta k_b$. This is shown in Fig.~\ref{ni62dp_dsde}(a) for the angle-integrated NEB differential cross section as a function of the proton energy in the c.m.\ frame. The shaded region  corresponds to negative energies of the neutron, that is, transfer to bound states. 
%the neutron-target breakup threshold. Proton energies above this limit correspond actually to neutron transfer to bound states. 
Although these contributions could be accounted for using the procedure of Ref.~\cite{Uda87}, they have not been considered here for simplicity.  It is seen that, as the bin width decreases, the results stabilize and for $\Delta k \approx 0.04$~fm$^{-1}$ they are well converged.

In Fig.~\ref{ni62dp_dsde}(b) we compare the converged post-form IAV calculation (thick solid line) with the prior calculation (dashed line), for the same observable. 
  The agreement between the prior and post calculations  is seen to be very satisfactory, with only small differences possibly due to numerical inaccuracies. This agreement corroborates  the post-prior equivalence at the numerical level. The choice of one or another representation becomes therefore a matter of numerical convenience. 
% same observable for the post and prior present the angle-integrated energy  differential cross section, as a function of the proton energy ($E_p$),  in the c.m.\ frame. The thick solid line is the  calculation using the \sout{\textcolor{red}{post-form}} IAV model and the
We show also in this figure the separate contributions of the prior form calculation (i.e., UT, NO and IN), 
 according to Eq.~(\ref{eq:post_prior}). It is seen that the full IAV calculation and the UT result (thin solid line) are in clear disagreement, as anticipated in the introduction. 
% thin solid line is  the result with the UT prior-form formula, i.e., the first term in  
%Eq.~(\ref{eq:post_prior}). It is seen that these results are in clear 
%disagreement, as anticipated in the introduction. The non-orthogonality 
%(NO) and interference (IN) terms are represented by the dotted and 
%dot-dashed lines. The sum of the three terms, i.e., UT+NO+IN, which 
%corresponds to the \textcolor{red}{\sout{full prior-form  expression} 
%IAV expression} (\ref{eq:post_prior}), is given by the dashed line. The agreement with the post-form result i 

In Fig.~\ref{ni62dp_dsde}(c) we compare the calculations with the  experimental 
data from Refs.~\cite{Kle81,Mas89}, corresponding to the double differential 
cross section as a function of the proton energy and for a proton 
detection angle of $\theta_p=20^\circ$ in the LAB frame. We note that, in this experiment, compound 
nucleus contributions were estimated and subtracted so the data should 
mainly correspond to the direct breakup modes considered here. The EBU 
contribution was calculated with the CDCC formalism, which goes beyond DWBA since it treats  
Coulomb and nuclear couplings to all orders. For the NEB part, we 
display the results obtained with the IAV and UT models. It is seen that 
the sum EBU+ NEB(UT), represented by the thin solid line, largely underpredicts the data. In contrast, the 
sum EBU + NEB(IAV) (thick solid line) reproduces reasonably well the magnitude and shape 
of the data, except for some underestimation at the smaller energies and 
some overestimation at the larger ones. We note that the low-energy tail 
will be mostly affected by the compound-nucleus subtraction and hence 
some uncertainty is expected at these energies. Our results are in contrast with those 
reported in Ref.~\cite{Mas89}, who found an overestimation of the IAV model.

%---------------------------------------------------------------------------------------------------
% 209Bi(6Li,aX)
%----------------------------------------------------------------------------------------------------
\begin{figure}[tb]
\begin{center}
% {\centering \resizebox*{0.85\columnwidth}{!}{\includegraphics{li6bi_dsde.eps}} \par}
\includegraphics[width=0.8\columnwidth]{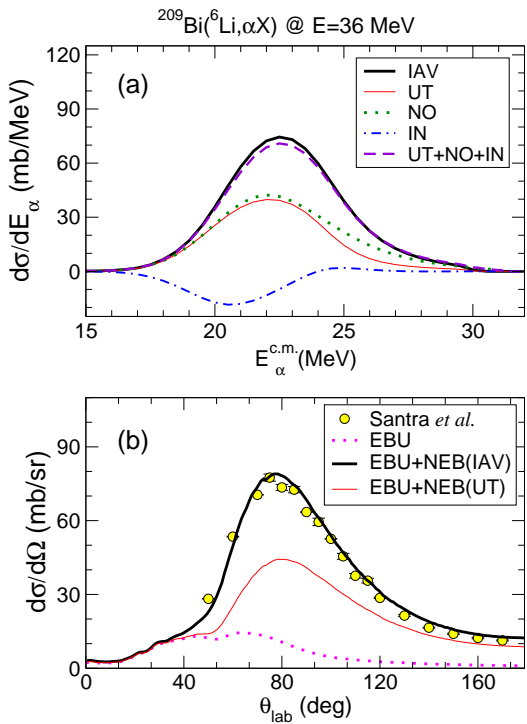}%
\caption{\label{li6bi_dsde}(Color online) (a) Angle-integrated energy 
differential cross section, as a function of the $\alpha$ c.m.\ enery, 
for the $^{209}$Bi($^6$Li,$\alpha$X) reaction at 36~MeV.  (b) Experimental and 
calculated angular distribution of  $\alpha$ particles, in laboratory 
frame, for the same reaction. The data are from Ref.~\cite{Santra11}.  }
\end{center}
\end{figure}
%-------------------------------------------------------------

As a second example, we consider the reaction  $^{209}$Bi($^6$Li,$\alpha$X), 
which  was also analyzed in our previous work \cite{Jin15}, using the post-form IAV model.  
 These calculations reproduced rather well the experimental angular 
distributions of $\alpha$ particles  for a wide range of  incident energies above and below the Coulomb 
barrier. To test the post-prior equivalence, we consider the incident energy 
of $E=36$~MeV. For the calculations presented here, we use the  potentials 
employed in  Ref.~\cite{Jin15}.

The results are shown in Fig.~\ref{li6bi_dsde}(a) for the angle-integrated 
$\alpha$ energy distribution (in the c.m. frame), with the same meaning 
for the lines as in Fig.~\ref{ni62dp_dsde}. The results are qualitatively 
similar to those found in the deuteron case, namely, (i) the post-form IAV 
model and the prior-form UT model yield significantly different results, and (ii) the sum UT+NO+IN gives a result very close to the post-form IAV model. Thus, the post-prior equivalence is also well fulfilled in this case.  
%and the addition of the NO and IN terms to the latter give a result very  close to the IAV result. 

In  Fig.~\ref{li6bi_dsde}(b) we compare these calculations with the data from Ref.~\cite{Santra11}, which correspond to  the angular distribution of $\alpha$ particles in the LAB frame. The EBU cross 
section corresponds to the CDCC calculation performed in Ref.~\cite{Jin15}, 
so we refer the reader to this reference for further details on this calculation.  
The EBU+NEB(IAV) calculation (thick solid line) reproduces remarkably  
well the shape and magnitude of the data. In contrast, the EBU+NEB(UT) 
calculation, represented by the thin solid line, clearly underestimates 
the data. This result reinforces the reliability of the IAV model.

%-----------------------------------------------------------------------------
\section{Summary and conclusions \label{sec:sum}}
%-----------------------------------------------------------------------------
In summary, we have addressed the problem of the post-prior equivalence in the calculation of NEB cross sections within the closed-form DWBA models proposed in the 1980s by Ichimura, Austern and Vincent \cite{Aus81,Kas82,Ich85} and by Udagawa and Tamura \cite{Uda81}. 

We have performed calculations for the  $^{62}$Ni($d$,$p$X)  and $^{209}$Bi($^6$Li,$\alpha$X) reactions at 25.5 and 36 MeV, respectively. In both cases, we find an excellent agreement between the post and prior expressions of the IAV model, confirming this equivalence at a numerical level. Moreover, the IAV model reproduces rather well the data in both reactions. In contrast, the UT model has been found to underestimate the experimental cross sections.  In the $^{62}$Ni($d$,$p$X) case, our results disagree with those of Ref.~\cite{Mas89}, which were used to criticize  the theory of IAV.  

The results presented in this work, along with those presented in related works \cite{Car15,Jin15,Pot15}, indicate that the IAV model provides a reliable framework to calculate NEB cross sections in reactions induced by deuteron and $^6$Li projectiles. Possible applications to other systems and problems are currently under study.

\begin{acknowledgments}
We are grateful to B.~Carlson and M.~G\'omez-Ramos for a critical reading of the manuscript. 
This work has been partially supported by the Spanish Ministerio de Econom\'ia y Competitividad, under grant  FIS2013-41994-P,  by the Spanish Consolider-Ingenio 2010 Programme CPAN
(CSD2007-00042)  and by Junta de  Andaluc\'ia (FQM160, P07-FQM-02894).
J.L.~is partially supported by a grant funded by the China  Scholarship Council. 
\end{acknowledgments}

% Bibliography
\bibliography{inclusive_prc.bib}
\end{document}